\documentclass{elsart}

\usepackage{natbib}
\usepackage{graphicx}

\usepackage{amssymb}

\begin{document}

\begin{frontmatter}



\title{Mode selection mechanism and magnetic diffusivity in a non-axisymmetric solar dynamo model}

 \author[]{Jie Jiang and Jingxiu Wang}
 \address {National Astronomical Observatories, Chinese Academy of
Sciences, Beijing 100012, China~~ E-mail:
jiangjie@ourstar.bao.ac.cn}

\begin{abstract}
The generation of solar non-axisymmetric magnetic fields is
studied based on a linear $\alpha^2-\Omega$ dynamo model in a
rotating spherical frame. The model consists of a solar-like
differential rotation, a magnetic diffusivity varied with depth,
and three types of $\alpha$-effects with different locations, i.e.
the tachocline, the whole convective zone and the sub-surface.
Some comparisons of the critical $\alpha$-values of axisymmetric
($m=0$) and longitude-dependent modes ($m=1,2,3$) are presented to
show the roles of the magnetic diffusivity in the problem of modes
selection. With the changing of diffusivity intensity for the
given solar differential rotation system, the dominant mode
possibly changes likewise and the stronger the diffusivity is, the
easier the non-axisymmetric modes are excited. The influence of
the diffusivity and differential rotation on the configurations of
the dominant modes are also presented.

\end{abstract}

\begin{keyword}
Sun: magnetic fields; Sun: diffusivity; Dynamo theory


\end{keyword}

\end{frontmatter}

\section{Introduction}
\label{} Since Carrington (1863) suspected that sunspots did not
form randomly over solar longitudes, many researches were
undertaken to investigate the preferred longitudes of solar
magnetic field. `Active Longitudes' \citep{sve69}, `Sunspot Nests'
\citep{cas86}, `Hot Spots' \citep{bai87} and so on were defined
and presented. The well-known `flip-flop' phenomena of late-type
stars were also identified on the Sun \citep{ber03}. These more
and more observational evidences indicate the involvement of
non-axisymmetric, i.e. longitude-dependent large-scale magnetic
fields in the formation and evolution of solar activities.

Although many papers reported the evidences for the existence of
longitudinal inhomogeneities in sunspots distribution, the results
were always inconsistent with each other on the number of the
active longitudes, i.e. longitudinal wave number $m$.
\citet{alt74} analyzed the large-scale photospheric magnetic field
in terms of surface harmonics and showed that the modes $m\leq 3$
were prominent throughout the solar cycle. \cite{det00} argued
that the total numbers of active regions per rotation per
hemisphere varied between 0 and 7. \cite{son05} suggested that  55
percent of the solar magnetic flux can be represented by a mode of
$m=6$ in the northern hemisphere and by $m=5$ in the southern one.
However, the identification of flip-flops on the Sun \citep{ber03}
indicates the mainly non-axisymmetric mode $m=1$. The ambiguous
longitudinal wave numbers may arouse a series of questions. Which
kinds of physical mechanisms are responsible for the preferred
mode to excite? For each mode, what is its configuration and which
ingredients have the primary influence on the configuration? These
are the main objectives of the paper.

In order to explain the questions about the non-axisymmetric field
of the Sun, it is natural to set up the non-axisymmetric solar
dynamo model. There are two basic processes in the traditional
$\alpha-\Omega$ dynamo model \citep{par55, ste66}: (i) the
generation of toroidal field by shearing a pre-existing poloidal
field (the $\Omega$-effect) and (ii) the re-generation of poloidal
field from the toroidal field (the $\alpha$-effect) . Moreover,
meridional circulation is always considered to transport the
magnetic flux. The $\alpha-\Omega$ dynamo is the limit of
$\alpha^2-\Omega$ dynamo in which the $\alpha$-effect is included
as a second generation source of the toroidal field. The role of
the $\alpha$-effect and the $\Omega$-effect in dynamo process has
been well-researched and well-known. But there is another
important ingredient, i.e. the magnetic diffusivity. Except for
opposing the generation of magnetic field, what other roles does
it play in the non-axisymmetric dynamo process?

Magnetic diffusivity consists of molecular diffusivity and
turbulent one. Since turbulent diffusivity is much stronger, we
neglect the molecular diffusivity. \citet{yos843} showed that
anisotropic turbulent diffusivity is a key factor on parity
selection especially for the latitudinal-gradient-dominated
differential rotation. \citet{cha04} adopted different
diffusivity-profiles (scalars) for toroidal and poloidal fields in
a two-dimensional solar dynamo model, based on the
helioseismically determined solar rotation profile and
Babcock-Leighton-type $\alpha$-effect. The solar-like dipolar
parity was obtained. Obviously, the diffusivity has directly
relation with the parity selection in dynamo process. In the
following, we also regard the diffusivity as a scalar and discuss
its influence on the mode selection and magnetic configuration
during the non-axisymmetric dynamo process.

In the next section, we set up a linear non-axisymmetric
$\alpha^2-\Omega$ dynamo model in a rotating frame with solar
inner core and introduce the spectral method to numerically solve
it. We present the role of turbulent diffusivity as well as the
differential rotation by the comparisons of the critical
$\alpha$-value of axisymmetric and longitude-dependent modes, and
further show their configurations in Section 3. In Section 4, We
give some discussion and conclusions.

 \section{The Non-axisymmetric Dynamo Model}

\subsection{The basic equations and numerical scheme}

We solve the standard mean field dynamo equation
\begin{equation}
\frac{\partial\textbf{\emph{B}}}{\partial\emph{t}}=\nabla\times[\emph{\textbf{U}}\times
\emph{\textbf{B}}+\alpha\textbf{\emph{B}}
-\eta\nabla\times\textbf{\emph{B}}],
\end{equation}
in the spherical polar coordinates $(r,\theta,\phi)$.  For the
flow field {\em \textbf{U}}, only the (differential) rotation
$\Omega$ without the meridional circulation is considered for
simplicity. $\eta$ is the magnetic diffusivity. We
non-dimensionalize the equation (1) in terms of the length
$R_\odot$ and the time $R_\odot^2/\eta_o$. Thus Eq. (1) becomes
\begin{equation}
\frac{\partial\textbf{\emph{B}}}{\partial\emph{t}}=R_\Omega\nabla\times(\emph{\textbf{U}}_\phi
'\times \emph{\textbf{B}})+R_\alpha\nabla\times(\alpha
'\textbf{\emph{B}}) -\nabla\times(\eta
'\nabla\times\textbf{\emph{B}}),
\end{equation}
where $R_\alpha=\displaystyle\frac{\alpha_o R_\odot}{\eta_o},
R_\Omega=\displaystyle\frac{|\Omega|_o R_\odot^2}{\eta_o}$.
$\eta_o$, $\alpha_o$ and $|\Omega|_o$ are the reference values of
the diffusivity in the convective zone (CZ), the $\alpha$-effect
and the differential rotation respectively. The quantities
$R_\alpha$ and $R_\Omega$ are the dynamo numbers measuring the
relative importance of inductive versus diffusive effects.

At two interfaces $r=r_o=1.0$ and $r=r_i=0.6$ (after
non-dimensionalization), both magnetic field and the tangential
electric field must be continuous. The exterior $r>1.0$ is a vacuum
and eigensolutions are matched to a potential field. The radiative
core is assumed to behave as a perfect conductor (See \citet{sch01},
\cite{jia06} for detail).

Since the magnetic field is divergence-free, we expand
$\textbf{\emph{B}}$ in term of two scalar functions \emph{h} and
\emph{g} which represent the poloidal and toroidal potentials as
\citep{mof78}
\begin{equation}
\textbf{\emph{B}}=\nabla\times\nabla\times\emph{\textbf{r}}\emph{h}
(r,\theta,\phi,t)+\nabla\times\textbf{\emph{r}}\emph{g}(r,\theta,\phi,t).
\end{equation}

With given boundary conditions and linear frame, we can look for
the eigensolutions of the equations about $h$ and $g$ with the
form
\begin{equation}
[\emph{h}(\emph{r},\theta,\phi,t),\emph{g}(\emph{r},\theta,\phi,t)]=
[\emph{h}(\emph{r},\theta,\phi),\emph{g}(\emph{r},\theta,\phi)]\mathrm{e}^{st},
\end{equation}
where \emph{s} is the eigenvalue and can be written as
$\emph{s}=\sigma+\mathrm{\emph{i}}\omega$. Only the solutions that
neither grow nor decay, i.e. the onset of dynamo actions
($\sigma=0$) are considered in the paper. For given $R_\Omega$,
the corresponding dynamo number $R_\alpha$ is the critical
$R_\alpha$ and the corresponding value of $\alpha_o$ is the
critical $\alpha$-value.

Different modes are decoupled in the linear theory. We expand $h$
and $g$ at the onset of the dynamo action in terms of Chebyshev
polynomial $T_{n}(r)$ and surface harmonics $P_l^m
\mathrm{e}^{\mathrm{i} m\phi}$ in the meridional circular sector
$r\in[0.6,1.0]$, $\theta\in[0.0,\pi]$ for given $m$ as follows:
\begin{eqnarray}
 h=\sum_{n=0}^{N}\sum_{l=m}^{L}c_{n,l}^h
 T_n(a r-b)P_l^m(\cos\theta)\mathrm{e}^{\mathrm{i}m\phi},\\
 g=\sum_{n=0}^{N}\sum_{l=m}^{L}c_{n,l}^g
 T_n(a r-b)P_l^m(\cos\theta)\mathrm{e}^{\mathrm{i}m\phi},
\end{eqnarray}
where $a r-b\in[-1,+1]$. $N$ and $L$ are the truncation levels to
get convergence. It varies with different dynamo numbers and
different $\Omega$, $\eta$, $\alpha$ profiles. $c_{n,l}^h$ and
$c_{n,l}^g$ are eigenvectors.

As pointed out by \citet{iva85}, the system of equations about $h$
and $g$ may be decomposed into two subsystems, i.e. odd or even
parity with respect to the equatorial plane denoted by $A$ and
$S$. With the parameters adopted in our model, both odd and even
parity solutions have nearly the same critical $R_\alpha$ and
actions for given mode $m$. It should be enough to present one
kind of parity for each mode. Since the odd parity for
axisymmetric mode $A0$ has been confirmed by the observations, we
will discuss it and the opposite parity (even ones) for the
non-axisymmetric modes ($Sm$). Note that, it does not mean there
have no possibility for $S0$ and $Am$ existing on the Sun. Through
studying the heliospheric magnetic field, \citet{mur04} gave the
evidence for $S0$ mode on the Sun. For the non-axisymmetric modes,
just like \citet{sti71}, only the modes $m=1, 2, 3~(S1, S2, S3)$
are considered since they are the most interesting ones in view of
observational evidences.

\begin{figure}
  \centering
\includegraphics[width=120mm]{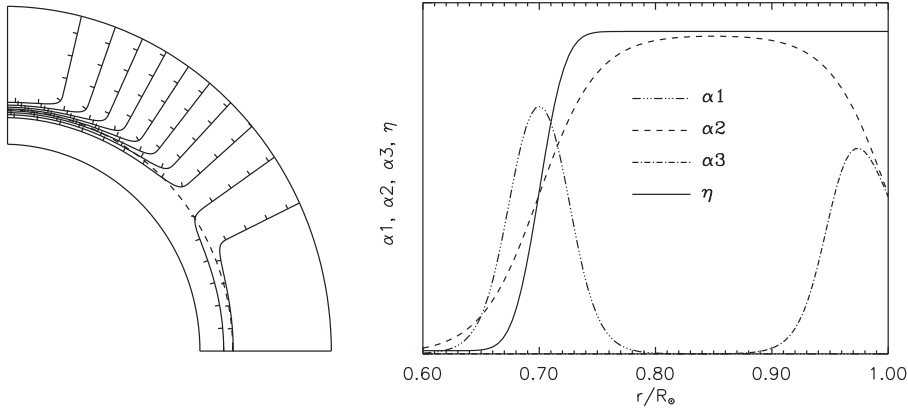}
\caption{Left: Contours of constant angular velocity obtained from
Eq. (7). The short, perpendicular ticks mark the `downhill'
direction. The dashed line denotes the center of the tachocline
locating at $0.7R_\odot$. The differential rotation is symmetric
about the equator and only quadrant is shown. Right: Radial
distributions of the three types of $\alpha$-effects and the
turbulent diffusivity $\eta$. Values of y-axis are not scaled.}
\end{figure}

\subsection{Three ingredients in the model}
We use the following expression for the solar internal rotation
\citep{sch98,cha99}
\begin{equation}
\Omega(r,\theta)=\Omega_c+\frac{1}{2}[1+erf(2\frac{r-r_c}{w})](\Omega_s(\theta)-\Omega_c),
\end{equation}
where
$\Omega_s(\theta)=\Omega_{EQ}+a_2\cos^{2}\theta+a_4\cos^{4}\theta$
is the surface latitudinal rotation. The parametric values are set
as $r_c=0.7R_\odot$, $w=0.05R_\odot$, $\Omega_c/2\pi=430.0~\rm
nHz$, $\Omega_{EQ}/2\pi=455.8~\rm nHz$, $a_2/2\pi=-51.2~\rm nHz$,
$a_4/2\pi=-84.0~\rm nHz$. It closely resembles the best-fit
helioseismic solution. Left of Fig. 1 is the contours of constant
angular velocity. Within the tachocline, rotation increases with
distance from the core at low latitudes, while it decreases at
high latitudes. At intermediate latitudes (near $35^\circ$),
rotation is almost independent on depth. Furthermore, we base our
model on the rotating spherical systems with the rotation velocity
$\Omega_c$ of the inner core. Thus the differential rotation in
the rotating frame $\Omega_c$ is
\begin{eqnarray}
\Omega^{'}(r,\theta)&=&(2\pi25.8)\times\frac{1}{2}[1+erf(2\frac{r-r_c}{w})]\nonumber\\
& &(1.-1.98\cos^{2}\theta-3.26\cos^{4}\theta)~ \rm{(nHz)}.
\end{eqnarray}

The surface equatorial differential rotation is ($2\pi25.8$) nHz
and we regard it as the characteristic value of the differential
rotation $|\Omega|_o$. Hence, $R_\Omega$ should be equal to
$\frac{\displaystyle 8\times10^{10}~\rm{m^2
s^{-1}}}{\displaystyle\eta_o}$ and be decided only by the
reference value of the diffusivity $\eta_o$.

We make the simple assumptions that both $\alpha$-effect and
magnetic diffusivity $\eta$ are scalars. Whilst bearing in mind
that they should be written as tensor quantities in a more
realistic model. We use the analytical expression of \citet{dik99}
for the diffusivity profile as
\begin{equation}
\eta(r)=\eta_c+\frac{\eta_o}{2}[1+erf(2\frac{r-r_c}{w})],
\end{equation}
which can be seen in the right of Fig. 1 (solid line). The
diffusivity $\eta_o$ in CZ, is dominated by its turbulent
contribution. In the stably stratified core, the diffusivity
$\eta_c$ is much lower because of the much less turbulence. In the
following, we take $\eta_c/\eta_o$=0.01. The transition from high
to low diffusivity occurs near the tachocline, which is coincident
with the rotational shear layer. Here, $\eta_o$ is far less
definite but is widely known that it ranges from $2\times
10^{10}~\rm cm^{2}s^{-1}$ to $2\times 10^{12}~\rm cm^{2}s^{-1}$.

The $\alpha$-effect cannot yet be determined from observations. We
consider three typical cases with different regions it works.
They are in the tachocline, the convective zone and the
near-surface. We denote them as C1, C2 and C3 respectively. The
expression for them can be written as
\begin{equation}
\alpha(r,\theta)=\alpha_o~\frac{1}{2}[1+erf(\frac{r-r_a}{d})]
~\frac{1}{2}[1-erf(\frac{r-r_b}{d})]\cos\theta.
\end{equation}
$r_a$, $r_b$ and $d$ change with different types of
$\alpha$-effects. $\alpha_o$ occurred in the dynamo number
$R_\alpha$ also presents the amplitude of the $\alpha$-effect. The
radial profiles are shown in the right of Fig. 1. The common
angular dependence $\cos\theta$ is adopted, which is the simplest
guaranteeing antisymmetry across the equator. Moreover, we do not
consider the $\alpha$-quenching since only the linear solutions
are sought.

 \section{Numerical results}
 \label{}
We perform some numerical explorations with the above three
different $\alpha$-profiles. In order to reveal the ingredients
contributing to the mode selection, we test $R_\Omega$ at the
values of 500, 2000 and 5000 and accordingly, the typical values
of diffusivity are $1.6\times 10^{12}~\rm cm^2 s^{-1}$, $4.0\times
10^{11}~\rm cm^2 s^{-1}$, $1.6\times 10^{11}~\rm cm^2 s^{-1}$.

\begin{table}
  \caption[]{The results of different modes for the case C1 with different $R_\Omega$ and
  $\eta_o (10^{11} \rm~cm^2 s^{-1})$.
  Middle three columns show critical $\alpha$-values ($\rm cm ~ s^{-1}$).
  The right three   columns are the corresponding period (years) and phase shift between $A0$ and $Sm$.
  Bold is used when the mode has the smallest $\alpha$-value.}
  \begin{center}\begin{tabular}{ccccccccc}
  \hline\hline\
~ & ~ & \multicolumn{3} {c} {Critical $\alpha$} &  ~ & \multicolumn{3} {c} {Period $T$ \& Phase shift}\\
\cline{3-5} \cline{7-9} $R_\Omega$ &~ & 500   & 2000
& 5000 & ~ & 500 & 2000 & 5000 \\
$\eta_o~ $ &~ & 16.0   & 4.0
& 1.6 & ~ & 16.0 & 4.0 & 1.6 \\
\hline\hline
$A0$ &~ & 104.8 & 9.91 & \textbf{1.6} & ~ & 1.85   & 6.8  & 16.88   \\
$S1$ &~ & 74.85 & \textbf{6.70}  & 2.1 & ~ & 12.18 & 286.9 & 7347.7 \\
~~~~ &~ & ~~~~~ & ~~~~~& ~~~~ & ~ &$0.84\pi$ & $0.69\pi$ & $1.04\pi$\\
$S2$ &~ & \textbf{73.60} & 10.77 & 4.2& ~ & 35.33 & 1207.6 & 3764.5 \\
~~~~ &~ & ~~~~~ & ~~~~~& ~~~~ & ~ &$1.30\pi$ & $0.16\pi$ & $1.33\pi$\\
$S3$ &~ & 82.56 & 15.55 & 6.4& ~ & 71.75 & 7409.4 & 1742.5\\
~~~~ &~ & ~~~~~ & ~~~~~& ~~~~ & ~ &$0.48\pi$ & $0.79\pi$ & $1.70\pi$\\
\hline\hline
  \end{tabular}\end{center}
\end{table}

\begin{table}
  \caption[]{The same as Tab. 1, but for Case 2.}
  \begin{center}\begin{tabular}{ccccccccc}
  \hline\hline\
~ & ~ & \multicolumn{3} {c} {Critical $\alpha$} & ~& \multicolumn{3} {c} {Period $T$ \& Phase shift}\\
\cline{3-5} \cline{7-9} $R_\Omega$ & ~ & 500 & 2000 & 5000 & ~ &
500 & 2000 & 5000\\
$\eta_o~ $ &~ & 16.0   & 4.0 & 1.6 & ~ & 16.0 & 4.0 & 1.6
\\\hline\hline
$A0$ &~ & 88.0 & \textbf{5.89} & \textbf{0.95} & ~  & 4.55  & 17.31 &  43.06 \\
$S1$ &~ & \textbf{79.39} & 8.50 & 2.53 & ~ & 26.06 & 349.5 &  2170.3 \\
~~~~ &~ & ~~~~~ & ~~~~~& ~~~~ & ~ &$0.72\pi$ & $0.73\pi$ & $0.74\pi$\\
$S2$ &~ & 83.47 & 12.97& 4.65 & ~ & 64.88 & 739.2 &  8024.5 \\
~~~~ &~ & ~~~~~ & ~~~~~& ~~~~ & ~ &$1.75\pi$ & $1.00\pi$ & $1.00\pi$\\
$S3$ &~ & 93.57 & 17.7 & 6.87 & ~ & 119.6 & 1524.6 & 22587.4 \\
~~~~ &~ & ~~~~~ & ~~~~~& ~~~~ & ~ &$1.05\pi$ & $0.95\pi$ & $0.42\pi$\\
\hline\hline
  \end{tabular}\end{center}
\end{table}

\begin{table}
  \caption[]{The same as Tab. 1, but for Case 3 and critical $\alpha$-values ($\rm m ~ s^{-1}$).}
  \begin{center}\begin{tabular}{ccccccccc}
  \hline\hline\
~ & ~ & \multicolumn{3} {c} {Critical $\alpha$} &  ~ &
\multicolumn{3} {c} {Period $T$ \& Phase shift}
\\ \cline{3-5} \cline{7-9}
$R_\Omega$ &~ & 500   & 2000 & 5000 & ~ & 500 & 2000 & 5000 \\
$\eta_o~ $ &~ & 16.0   & 4.0
& 1.6 & ~ & 16.0 & 4.0 & 1.6 \\
\hline\hline
$A0$ &~ & 26.17 & \textbf{5.99} & \textbf{1.63} & ~ & 5.79 & 1.16 & 2.0 \\
$S1$ &~ & \textbf{16.77} & 4.55 & 1.88 & ~ & 0.297 & 0.293 & 0.292\\
~~~~ &~ & ~~~~~ & ~~~~~& ~~~~ & ~ &$0.10\pi$ & $1.10\pi$ & $0.7\pi$\\
$S2$ &~ & 17.84 & 4.99 & 2.11 & ~ & 0.150 & 0.147 & 0.146\\
~~~~ &~ & ~~~~~ & ~~~~~& ~~~~ & ~ &$0.10\pi$ & $0.10\pi$ & $1.29\pi$\\
$S3$ &~ & 19.12 & 5.46 & 2.22 & ~ & 0.101 & 0.100 & 0.099\\
~~~~ &~ & ~~~~~ & ~~~~~& ~~~~ & ~ &$0.72\pi$ & $0.25\pi$ & $1.27\pi$\\
 \hline\hline
  \end{tabular}\end{center}
\end{table}

\subsection{The comparisons of the critical $\alpha$-values among the different modes}
The middle three columns of Table 1-3 show the critical
$\alpha$-values for different $R_\Omega$ and $\eta_o$ of the case
C1 that the $\alpha$-effect concentrates in the tachocline
($r_a$=0.675, $r_b$=0.725 and $d$=0.025), C2 that the
$\alpha$-effect exists through the entire CZ ($r_a$=0.7, $r_b$=1.0
and $d$=0.05) and C3 that the $\alpha$-effect locates at the top
of the surface ($r_a$=0.95, $r_b$=1.0 and $d$=0.025),
respectively. The modes with the lowest critical $\alpha$-values
are shown in bold.

The general rule that the larger $m$ is, the more difficult the
mode to excite is only satisfied when $R_\Omega$ is large enough
and the corresponding diffusivity is lower enough. For example,
when $R_\Omega=5000$ ($\eta_o=1.6\times 10^{11} \rm~cm^2 s^{-1}$)
for all of the three cases and $R_\Omega=2000$ ($\eta_o=4\times
10^{11}~\rm cm^2 s^{-1}$) for C2 and C3, the models will favor the
axisymmetric mode $A0$. However, when the turbulent diffusivity is
strong ($\eta_o=1.6\times 10^{12}~\rm cm^2 s^{-1}$ for the three
cases), the non-axisymmetric modes ($m=2$ for C1, $m=1$ for C2 and
C3) have the lowest critical $\alpha$-values and will be the
preferred modes. Hence, the dynamo number $R_\Omega$ is the
decisive parameter to select the dominant mode. According to
$R_\Omega=\displaystyle\frac{|\Omega|_o R_\odot^2}{\eta_o}$, since
the differential rotation is given, only the diffusivity in CZ
contributes to the result  and hence should be the key factor to
the mode selection.

When $R_\Omega$ is large enough ($R_\Omega \gg R_\alpha$), the
$\alpha$-effect as the generation mechanism of toroidal field can
be ignored and the $\alpha^2-\Omega$ model is at the
$\alpha-\Omega$ dynamo limit. At the $\alpha-\Omega$ limit, the
axisymmetric mode ($m=0$) which is the smallest wave number and
has the largest scale field is least depressed by turbulent
diffusivity and will be the preferred mode. When $\eta_o$
increases and the corresponding $R_\Omega$ decreases, the ratio of
$R_\Omega$ to $R_\alpha$ also decreases. The role of
$\alpha$-effect in generation of the toroidal field increases and
the $\alpha^2-\Omega$ model deviates the $\alpha-\Omega$ dynamo
limit gradually. The more it deviates from the limit, the easier
the non-axisymmetric modes are excited, which is in conformity
with the asymptotic solution of \citet{bas05}.

The right three columns of Table 1-3 show the periods and phase
shifts for different modes with different $R_\Omega$ and $\eta_o$.
For the axisymmetric mode $A0$ of the cases C1 and C2, it can be
seen from the Table 1 and 2 that the period is nearly the inverse
ratio of diffusivity and hence, is dominated by the diffusivity.
The periods of the non-axisymmetric modes are much larger than
that of $A0$. Therefore they are weakly oscillating or rather
steady comparing with $A0$. This is consistent with observations
\citep{ber03} and theory speculation \citep{flu04}. The
co-existing of the two kinds of modes are used to explain active
longitudes, flip-flops and other non-axisymmetric phenomena
\citep{mos04,flu04}. In contrast, the oscillatory period of case
C3 is much less than the other two cases. This is consistent with
the independent results of \citep{iva85,big04}. The periods depend
on the diffusivity much less especially for the non-axisymmetric
modes. C3 is characterized by an $\alpha$-effect which has no
overlap with the shear layer. It probably is the reason to cause
the difference. The meridional circulation neglected in our models
should play much more important roles in C3 to determine the
cycle.

Phase relation between the surface axisymmetric and
non-axisymmetric poloidal field can be observed. Ruzmaikin et al.
(2001) showed that there are about $\pi$-shift between the modes
$A0$ and $S1$ and the nearly same phase between $S1$ and $S2$. The
global phases \citep{sch95}, based on the maximum values of $B_r$
over the whole latitude ranges are computed. The phase shifts
presented in Tab. 1-3 are the phase difference between $A0$ and
$Sm$. The results dependent on $R_\Omega$ ($\eta_o$) and different
cases. When $R_\Omega$ is large enough and diffusion is small, the
phase shift can match the observation well.

\begin{figure}
  \centering
\includegraphics[width=120mm]{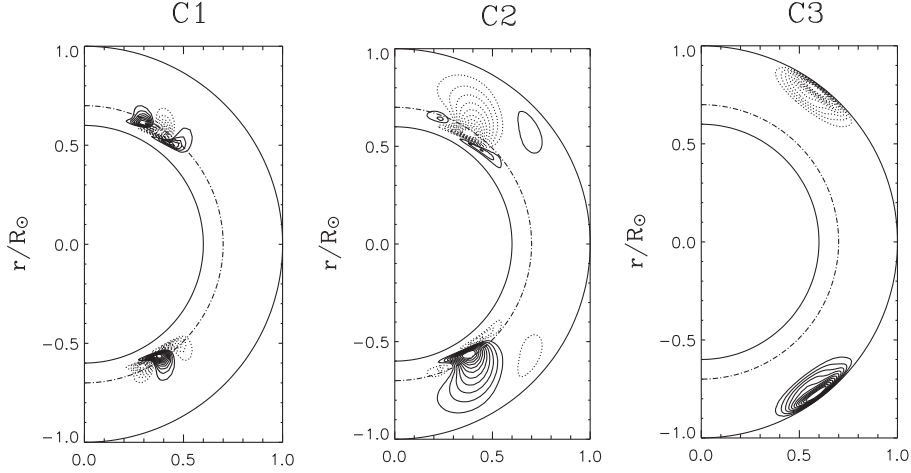}
\caption{Contours of the toroidal field $B_\phi$ of the axisymmetric
mode $A0$ in a meridional plane with $R_\Omega$=5000 for the three
cases with different locations of $\alpha$-effects. Solid (dashed)
contours correspond to positive (negative) magnetic field
components. The dot-dashed lines locate at $0.7R_\odot$ marking the
center of the tachocline. The radial distributions of the contour
lines are consistent with that of the $\alpha$-effects. But the
latitudinal locations which are around $55^\circ$ latitude do not
change with the $\alpha$-profiles.}
\end{figure}

\subsection{The configurations of the magnetic fields for the three cases}
Fig. 2 shows the contours of the toroidal field $B_\phi$ of the
axisymmetric mode $A0$ in a meridional plane for the three cases
when $R_\Omega$ is 5000. The centers of the contour lines
concentrate at the nearly same high latitudes (about $55^\circ$)
where have the strong radial shear. But the radial locations of
the three cases are different. They are around the tachocline,
throughout the whole CZ and near the surface respectively, which
are consistent with the radial profile of the $\alpha$-effects.
Hence, the strong radial shear of the differential rotation and
the $\alpha$-effect play a dominant role in the form of the
configuration of the axisymmetric fields.

Fig. 3 presents the distribution of $B_\phi$ of the distinguished
non-axisymmetric modes for the three cases when $R_\Omega$ is 500.
For C1 and C2, the magnetic field is highly concentrated in the
lower part of the tachocline, where the diffusivity is small (see
Eq. (9)). But it is different for C3 with the sub-surface
$\alpha$-effect. Moreover, for all the three cases, the
non-axisymmetric fields are localized around $35^\circ$ latitude
where has the weakest radial shear (see left of Fig. 1).

The physical reason for the different localizations of the
axisymmetric and non-axisymmetric modes can be understood from MHD
theory \citep{mof78}. The differential rotation affects the two
kinds of fields in different ways. It distorts the axisymmetric
poloidal field perpendicular to the axis of rotation to form the
axisymmetric toroidal field. By the way, this needs the non-zero
diffusivity to assure the development of net toroidal flux. For
the non-axisymmetric ingredients, they will expelled from the
strong differential rotating high-diffusivity region.

\begin{figure}
  \centering
\includegraphics[width=120mm]{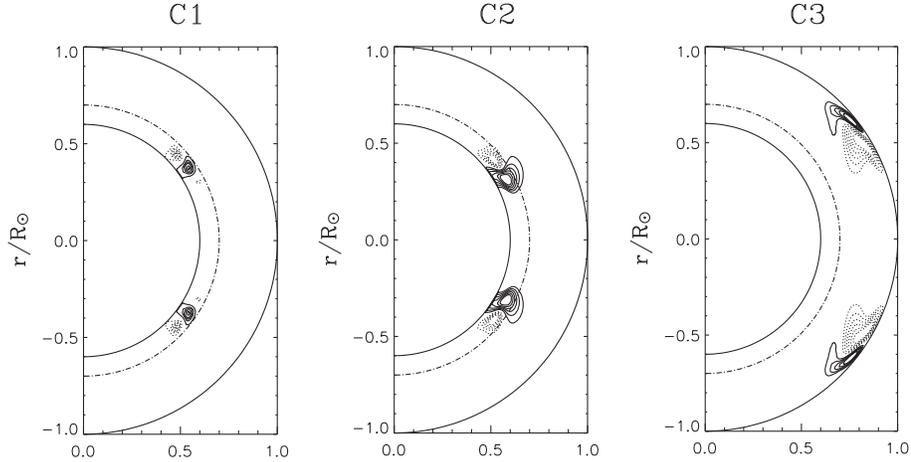}
\caption{Contours of the toroidal field $B_\phi$ of the
non-axisymmetric modes in a meridional plane with $R_\Omega$=500 for
the three different cases. The line styles are the same as in Fig 2.
For C1 and C2, the field concentrates in the low diffusivity (below
the dot-dashed line $0.7R_\odot$) and weak differential rotation
(about $35^\circ$ latitude). For C3, it has the same latitudinal
location but concentrates near the surface.}
\end{figure}

\section{Conclusions and Discussion}
 \label{}

By the comparisons of critical $\alpha$-values among the modes
$m\leq 3$ for a wide range of turbulent diffusivity and dynamo
number $R_\Omega$ accordingly, we obtain that dynamo number
$R_\Omega$ is the decisive parameter for the modes selection. When
$R_\Omega$ is large enough, the dominant axisymmetric mode would
be favored. With the decreasing of $R_\Omega$, different
non-axisymmetric modes will be preferred. Since $R_\Omega$ is the
ratio of the differential rotation to magnetic diffusivity, both
of the two ingredients should contribute to the modes selection.
The role of differential rotation and its physical mechanisms have
been well-known that the strong value prefers the axisymmetric
mode. For the definite differential rotation given by
helioseismology, the comparisons approve of the effect of the
diffusivity in the mode selection and indicate that it is just
contrary to the differential rotation. In other words, the strong
turbulent diffusivity favors the non-axisymmetric modes for the
given rotation system of the Sun.  On the other hand, the dominant
mode given by the observation can be an important index to infer
the inner turbulent diffusivity of the Sun or a star.   What we
have done is based on a linear non-axisymmetric $\alpha^2-\Omega$
dynamo model in a rotating spherical frame which incorporates the
solar-like differential rotation, the magnetic diffusivity varied
with depth and three possible cases of the $\alpha$-effects.

On the actual Sun, it is rather than a single preferred mode. The
whole physics during the dynamo process is not only the
interaction between magnetic field and turbulent velocity field
but also the interaction between the various modes. The latter may
lead to a stationary field amplitude and to a certain spectral
distribution, even if the $\alpha$-value is supercritical
\citep{sti71}. The nonlinearities, such as a non-axisymmetric
distribution of the $\alpha$-effect \citep{mos02, big04} and the
$\alpha$-quenching \citep{mos99}, induce the different modes
coupled together. The non-axisymmetric enhancement of the
underlying magnetic field causes in the clustering of sunspots to
form active longitudes \citep{ruz01a}.

Since the magnetic diffusivity is a diffusion of a vector field,
the diffusivity profile should be expressed by a tensor. Yoshimura
et al. (1984a ,1984b, 1984c) adopted the tensors for the
diffusivity and demonstrated that it had close relations with
parity selection. With a scalar expression, the growth rates for
the odd and even parity modes are nearly the same (Seenbeck and
Krause 1969; Deinzer and Stix 1971) which also been demonstrated
by our code. But in the paper, the scalar diffusivity is still
used for two reasons. One is for the simplicity. We do not refer
to the parity problem and only adopt the odd parity for
axisymmetric mode and even parity for the non-axisymmetric modes.
The other is limitation of the observations. The measurements of
magnetic diffusivity inside of the Sun are not possible nowadays.
Hence it is difficult to give its detailed distribution.

The configurations of the toroidal fields for the axisymmetric and
non-axisymmetric modes are different. The former prefers to locate
at the strong radial shear region and the latter favors the weak
radial shear and low diffusivity regions. These are caused by the
different roles that the differential rotation and diffusivity
played in the two kinds of modes. Furthermore, the meridional
circulation is not considered in the work. It plays an important
role in the axisymmetric mode \citep{nan02}. It carries the strong
axisymmetric toroidal field produced at the high latitudes to the
low ones and produces the active regions there with the magnetic
buoyancy. We will discuss its influence on the non-axisymmetric
field in the coming papers.


\section*{Acknowledgments}

We thank the reviewers, Arnab Rai Choudhuri and another anonymous
one, for useful comments that resulted in improvements of our paper.
We are grateful to Liao, X.H. for his careful supervision on the
development of the numerical code. This work has been supported by
National Natural Science Foundation of China (10573025, 10603008)
and by the National Key Basic Research Science Foundation
(G2006CB806303).

\end{document}